**Częstochowski**

**Kalendarz Astronomiczny**

**2013**

**Rok IX**

**Redakcja**

**Bogdan Wszołek**
**Agnieszka Kuźmicz**

**Wersja elektroniczna kalendarza jest dostępna na stronach**

**www.astronomianova.org**
**www.ptma.ajd.czest.pl**

# Special Shapes for Optimal Phenomenological Approximations of Astronomical Signals

Ivan L. Andronov

Department "High and Applied Mathematics", Odessa National Maritime University, Ukraine, tt_ari@ukr.net

We discuss different methods for phenomenological approximations of signals with (generally) irregularly spaced arguments. Such signals may be classified as periodic, multi-periodic, quasi-periodic (cyclic), burst-type and flicker-type (see e.g. a review by Andronov (2003)). In this talk, we concentrate on (nearly) periodic light curves of the Algol-type eclipsing variables. Having no possibility to determine all physical parameters for all stars and to make physical modeling, the phenomenological fits are used. This is especially important for discovery papers, when there are often available only mono-chromatic photometric data, without estimates of temperature and orbital velocities of the components.

Obviously, the physical modeling is better. However, it needs to determine more than a dozen of unknown parameters, and so generally, the signal may be fitted by a trigonometric polynomial (sometimes called "a multi-harmonic approximation") with a statistically optimal degree $s$. For determination of its value, some criteria may be used (Fischer's, Student's, of mean squared accuracy of the fit etc.). The set of corresponding complementary algorithms and programs was initially presented by Andronov (1994) and in further papers on newer methods.

For symmetric curves, we also use "symmetric" trigonometric polynomial fits with cosines without sines. In this case, an additional parameter - the optimal phase shift is determined.

However, for the curves with parts of abrupt changes (Algol-type variables, RR Lyrae - type stars), the number of determined parameters $m$ is large, which causes unrealistic waves at the fit. This effect is especially large in noisy signals.

To decrease $m$ for noisy signals, special shapes with a smaller number of parameters are to be used, increasing the "signal-to-noise" ratio. We introduce some special shapes like "splines of changing order" for RR Lyr-type variability, when the phase curve is approximated by a cubic parabola (short ascending branch) and a square parabola (descending branch). After non-linear fitting, all phenomenological parameters needed for the "General Catalogue of Variable Stars" are determined. We called this method "RR catcher", which was effectively used to discover faint variables from the Hipparcos-Tycho catalogue.

To determine characteristics of extrema needed for analysis of possible period variations, the "Asymptotic parabola" method is used.

Another "EA catcher" was used for eclipsing binaries, the effectivity of which increases with decreasing width of minimum. It was based on a polynomial spline approximation – a constant outside eclipses and parabolic minima of the same width and different depth at minima. This was also used for the variables discovered using the Hipparcos-Tycho.

However, for the EA, EB and even some EW-type binaries, we have proposed a more efficient "NAV" ("New Algol Variable") algorithm, when the light curve is approximated "semi-physically": trigonometric polynomial fit of order $s=2$ corresponding to effects of ellipticity and reflection and local additions at phases of primary and secondary minima. We have tried few shapes for best phenomenological modeling of the light curve.

The preliminary introduction of the method was presented by Andronov (2010), with a detailed description in a full-length paper by Andronov (2012). Results of the first application of our new method were published by Virnina (2010).

In the following figures, we show test functions and approximations with comments in figure captions "in a style of presentation" without repeating in the text. The figures and illustrative stars are different from that used in the Papers I and II. Thus this paper may be called as "Paper III".

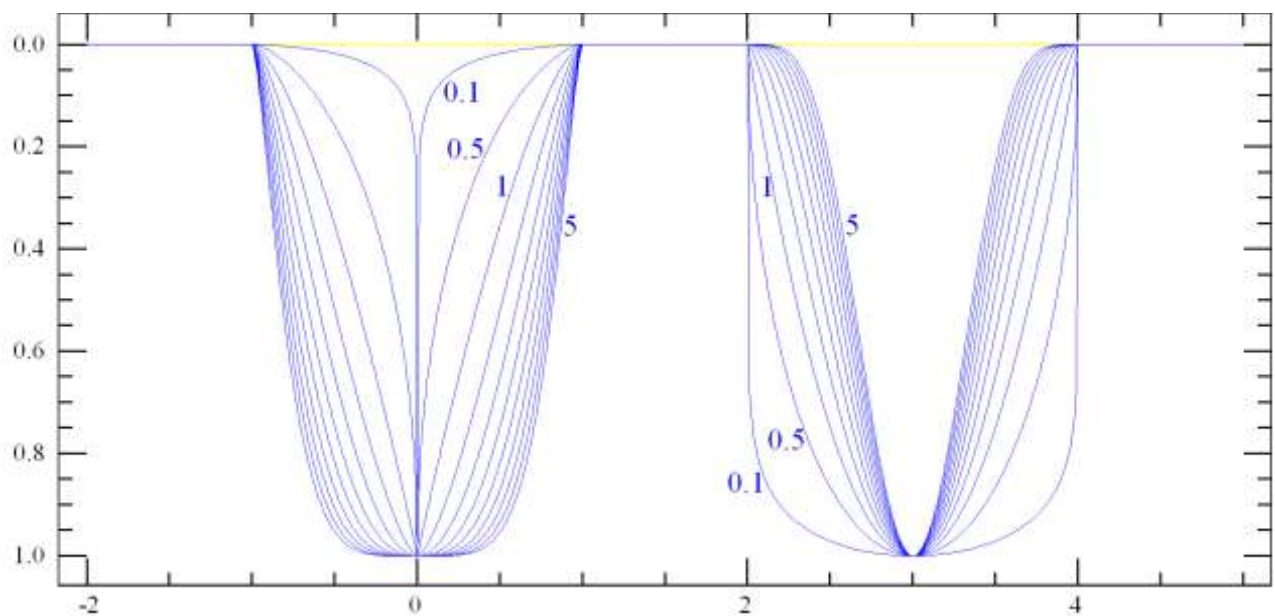

Fig.1. Families of additional shapes for minima of eclipsing variables: $H(t)=(1-|t|^\beta)^{3/2}$ ("NAV", left) and $H(t)=(1-t^2)^\beta$ (right) introduced by Andronov (2010, 2012), for the values of the parameter $\beta=0.1; 0.5; 1; 1.5; 2; 2.5; 3; 3.5; 4; 4.5; 5$. Both families are local, i.e. having finite width. However, the NAV functions correspond to a shape of eclipse better.

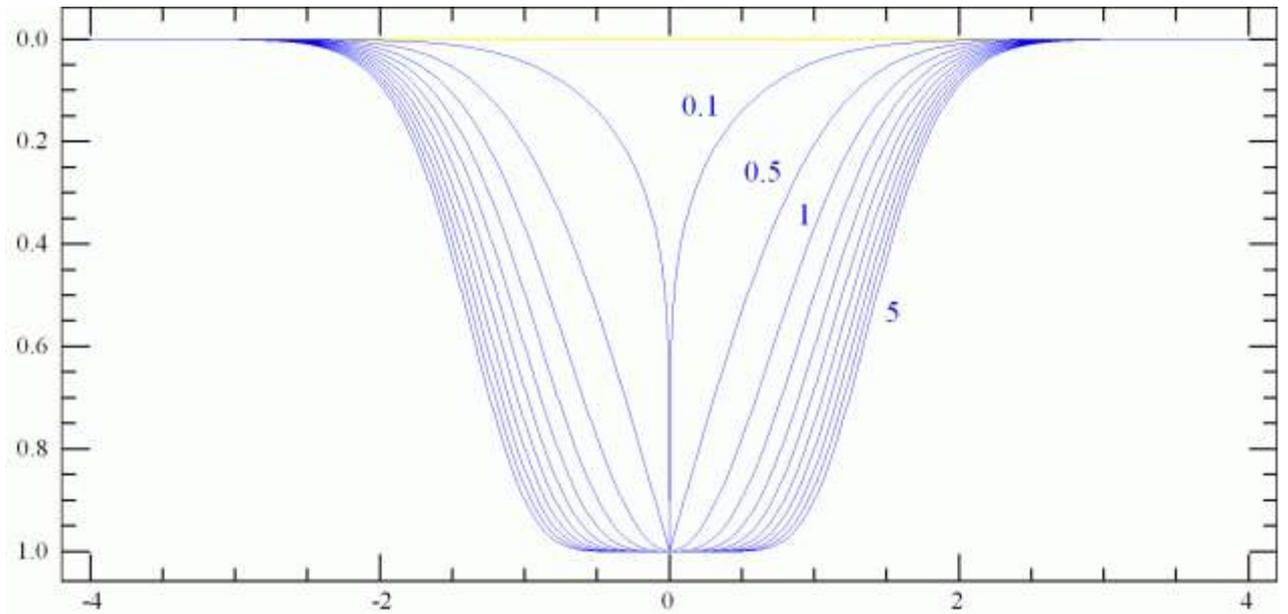

Fig.2. Families of additional shapes for minima of eclipsing variables: $H(t)=1-(1-\exp(-t^2))^\beta$ introduced by Mikulasek et al. (2012) and comparatively studied by Andronov (2012), for the values of the parameter $\beta$=0.1; 0.5; 1; 1.5; 2; 2.5; 3; 3.5; 4; 4.5; 5. Although near the center of the eclipse the shapes are similar to that shown in Fig.1, there are no definite borders $\beta$, thus the fit agrees worse with a physical model with well defined begin and end of the eclipse. Thus the NAV functions are more effective.

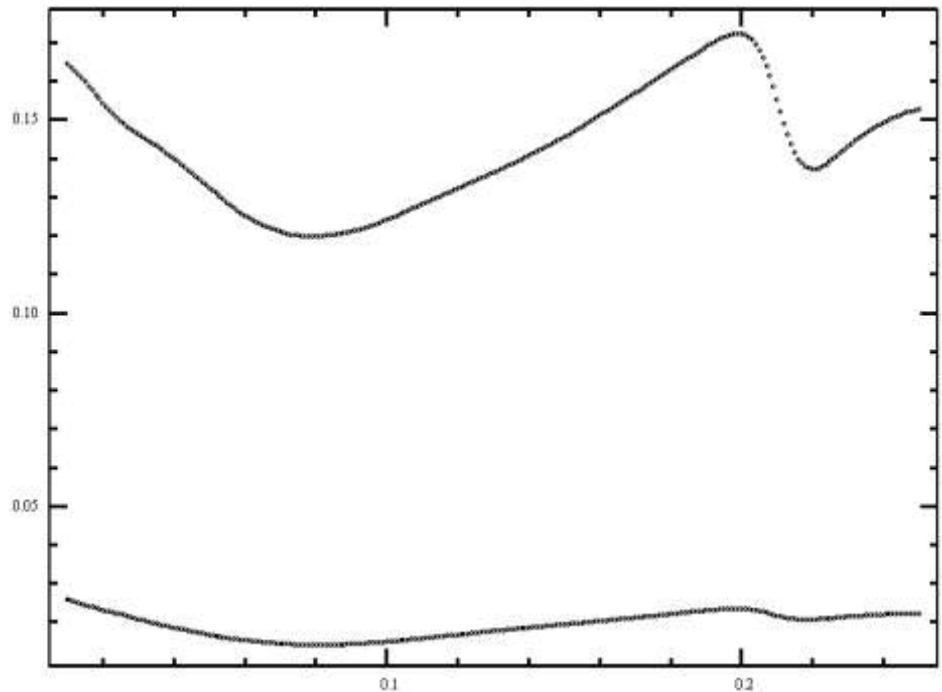

Fig.3. Dependence on the eclipse half-duration of the test function (unbiased r.m.s. deviation of the points from the fit). Most deep minimum corresponds to a statistically optimal value.

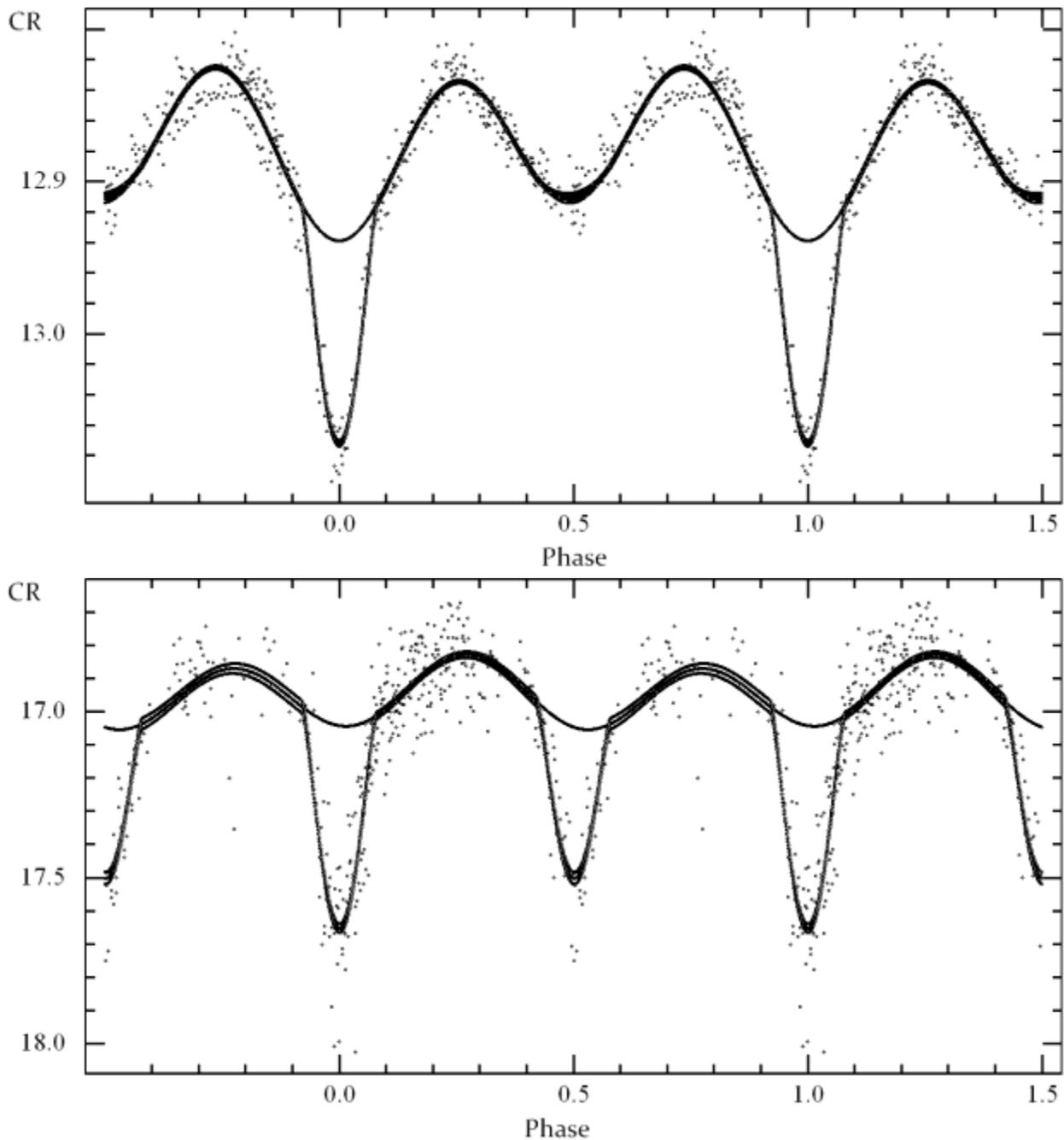

Fig.4. The phase light curves of two new Algol-type variables USNO-B1.0 1238-0228470 = VSX J141509.2+335222 and USNO-B1.0 1229-0276915 = VSX J141340.5+325648 in Boo, which were discovered by Virnina (2010) and its best fit using the NAV algorithm with corresponding "±1σ" corridor for the approximation. At the phases of eclipses also is shown an extension of the second-order trigonometric polynomial fit, which corresponds to effects of "reflection", "ellipticity" and "spots" (O'Connell).

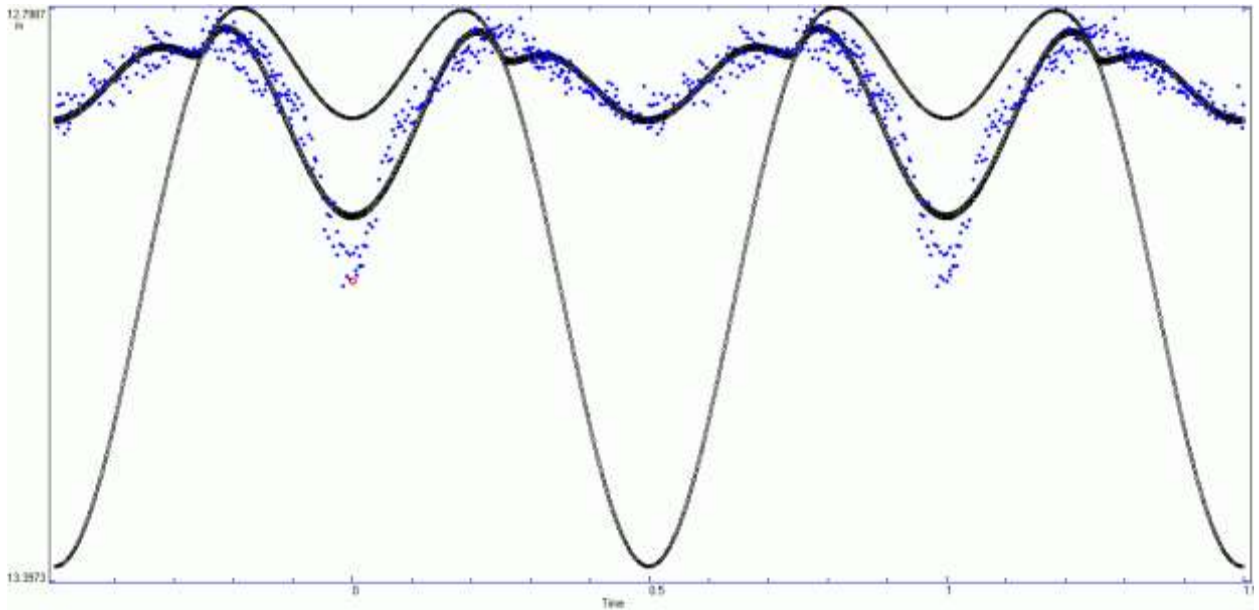

Fig.5. The New Algol-type variable USNO-B1.0 1238-0228470 = VSX J141509.2+335222 and USNO-B1.0 1229-0276915 = VSX J141340.5+325648 in Boo discovered by Virnina (2010) and its approximation using a" wrong" half-width D/2=0.25P, which corresponds to a second local minimum at this value. Although the formal fit (thick line) exists and is even not very bad, it has no physical sense, as one may see from the "out-of eclipse" curve (thin line), because: 1) no eclipses may be as wide as half of the period; 2) at the secondary minimum, the resulting curve is "brighter" than it's "out-of-eclipse" part, what challenges the hypothesis of eclipses. This figure is shown for an illustration of justification of a single solution at the global minimum of the test function (Fig. 3). The screenshot is from the program "MCV" (Andronov and Baklanov, 2004).

Our methods for time series analysis have been applied to 1400+ variable stars of different types in a course of the "Inter-Longitude Astronomy" (Andronov et al. 2010) and "Ukrainian Virtual Observatory" (Vavilova et al. 2012) projects.

\*\*\*

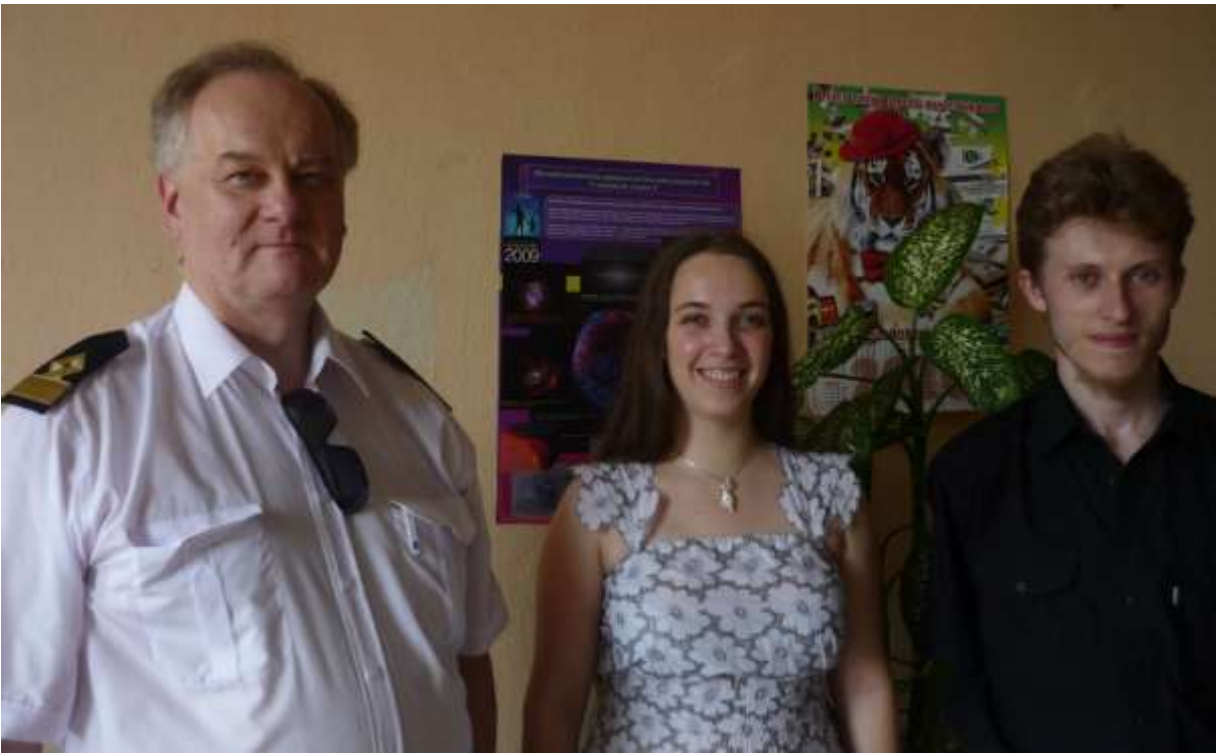

Prof. Ivan L. Andronov and his "scientific children":
Natalia A. Virnina and Vitalii V. Breus. (*fot. I.Andronov*)